\documentclass{ws-procs9x6}
\usepackage{epsfig}   
\begin{document}

\title{Extended Cluster Model for Light, and Medium Nuclei}

\author{M. Tomaselli$^{*}$, T. K\"uhl, D. Ursescu, and S.Fritzsche
\address{GSI, Gesellschaft f\"ur Schwerionenforschung, Darmstadt\\
D24691 Germany}
$^*$E-mail: m.tomaselli@gsi.de}

\begin{abstract}
The structures, the electromagnetic transitions, and the beta decay strengths of exotic nuclei
are investigated within an extended cluster model. 
We start by deriving an effective nuclear Hamiltonian within the $S_2$ correlation operator.
Tensor forces are introduced in a perturbative expansion which includes up to the
second order terms. Within this Hamiltonian we calculate the distributions and the 
radii of A=3,~4 nuclei.
For exotic nuclei characterized by n valence protons/neutrons we excite
the structure of the closed shell nuclei via mixed modes formed by
considering correlations operators of higher order.
Good results have been obtained for the calculated transitions and for the beta decay
transition probabilities.

\end{abstract}


\bodymatter

\section{Introduction}\label{aba:sec1}
The data obtained at RIA, Riken and GSI have given new life to the old Nuclear Physics.
Nuclear structures of proton and neutron rich nuclei are and will be investigated giving a new
insight to the fundamental observable such as the nuclear forces in the proton-proton
and neutron-neutron components, shell closure far from stability, magnetic properties of weakly
excited nuclear states, and many others.  
The theoretical analysis of these data requires  
a reliable nuclear model which can reproduce the data of stable nuclei and be extrapolated to predict
or at least reproduce the experimental results.  
The Cluster Correlation Model offers effective tools in this direction. 
We have to depart, however, from a perturbative scheme which is generally used to treat the two body correlation.
Generalization to the perturbative method to include the many body correlation is in this paper realized 
within an extended, non-perturbative Cluster Correlation Model.
One of the central challenges of theoretical nuclear physics is the attempt to
describe unknown properties of the exotic systems in terms of a realistic
nucleon-nucleon (NN) interaction.
In order to calculate matrix elements with the singular interaction (hard core)
we have to define effective correlated Hamiltonians. 

Correlation effects in nuclei have been first introduced in nuclei by Villars~\cite{vil63},
who proposed the unitary-model operator (UMO) to construct effective operators.
The method was implemented by Shakin~\cite{sha66} for the calculation
of the G-matrix from hard-core interactions.
Non perturbative approximations of the UMO have been recently applied to odd nuclei 
in Ref.~[3] and to even nuclei in Ref.~[4]. 
The basic formulas of the Dynamic Correlation Model and of the
 Boson Dynamic Correlation Model (BDCM) presented in the above
quoted papers have been obtained by separating the n-body correlation 
operator in short- and long-range components.
 The short-range component is considered up to the two body correlation
while for the long range component 
the three and four body correlation operators have been studied.
The extension of the correlation operator to high order diagrams is especially important
in the description of exotic nuclei (open shell).  
In the short range approximation 
the model space of two interacting particles is separated in two subspaces:
one which includes the shell model states and the other (high momentum) which
 is used to compute the G-matrix of the model.
 The long range component of the correlation operator 
 has the effect of generating a new correlated model space (effective space)
which departs from the originally adopted one (shell model).
The amplitudes of the model wave functions are calculated in terms of
non linear equation of motions (EoM).
 The derived systems of commutator equations,
which characterize the EoM, are finally linearized.
Within these generalized linearization approximations (GLA)
we include in the calculation presented in the paper up to the ((n+1)p1h) effective diagrams.
The linearized terms provide, as explained later in the text, the 
additional matrix elements that convert the perturbative UMO expansion 
in an eigenvalue equation. The n-body matrix elements needed to diagonalize the 
 resulting eigenvalue equations are calculated exactly 
via the Cluster Factorization Theory (CFT)~\cite{tom06}. \\ 

\section{The $S_2$ correlated Hamiltonian}
In order to describe the structures and the distributions of nuclei
we start from the following Hamiltonian:
\begin{equation}\label{equ.1}
H= \sum_{\alpha\beta}
 \langle\alpha|t|\beta\rangle \, a^{\dagger}_{\alpha}a_{\beta}
   \:+\: \sum_{\alpha\beta\gamma\delta}
   \langle\Phi_{\alpha\beta}| v_{12}|\Phi_{\gamma\delta}\rangle \,
   a^{\dagger}_{\alpha}a^{\dagger}_{\beta} a_{\delta}a_{\gamma}  
\end{equation}
where $v_{12}$ is the singular nucleon-nucleon two body potential.
Since the two body states $|\alpha\beta\rangle$ are uncorrelated the matrix elements of 
$v_{12}$ are infinite. This problem can be avoided by taking matrix elements
of the Hamiltonian between correlated states.
In this paper the effect of correlation is introduced via the $e^{iS}$ method.
In dealing with very short range correlations only the $S_2$ part of the correlation
 operator needs to be considered.  

Following Ref.~[2] we therefore calculate an ``effective Hamiltonian'' by using only the $S_2$ 
correlation operator obtaining:
\begin{equation}\label{eq.1}
\begin{array}{l}
H_{eff}=e^{-iS_2}He^{iS_2}=\sum_{\alpha\beta}\langle\alpha|t|\beta\rangle a^{\dagger}_{\alpha}
a_{\beta}+
\sum_{\alpha\beta\gamma\delta}\langle\Psi_{\alpha\beta}|v^l_{12}|\Psi_{\gamma\delta}\rangle 
a^{\dagger}_{\alpha}a^{\dagger}_{\beta} a_{\delta}a_{\gamma}\\
=\sum_{\alpha\beta}\langle\alpha|t|\beta\rangle a^{\dagger}_{\alpha}a_{\beta}+
\sum_{\alpha\beta\gamma\delta}\langle\Psi_{\alpha\beta}|v|\Psi_{\gamma\delta}\rangle a^{\dagger}_{\alpha}a^{\dagger}_{\beta} a_{\delta}a_{\gamma}
\end{array}
\end{equation}
where $v_{12}^l$ refers to the long-range part of the nucleon-nucleon force
diagonal in the relative orbital angular momentum,
after the separation~\cite{mos60}:
\begin{equation}
\label{eq.2}
v_{12}=v_{12}^s+v_{12}^l 
\end{equation}
The separation is made in such a way that the short range part produces no
energy shift in the pair state~\cite{mos60}.
In doing shell model calculation with the Hamiltonian Eq.~(\ref{eq.1}),
we remark: a) only the long tail potential plays an essential role in the calculations
of the nuclear structure i.e.: the separation method and the new
proposed $v_{low-k}$~\cite{jia01} method show a strong analogy and
b) the $v_T^{od}$ must be included as an additional re-normalization of the
effective Hamiltonian Eq.~(\ref{eq.1}).
 
In Eq.~(\ref{eq.1}) the $\Psi_{\alpha\beta}$ is the two particle correlated wave function:
\begin{equation}\label{eq.1a}
\Psi_{\alpha\beta}=e^{iS_2}\Phi_{\alpha\beta}
\end{equation}

In order to evaluate the effect of the tensor force on the $\Psi_{\alpha\beta}$
 we calculate:
\begin{equation}
\label{eq.3}
w(r)=V_T^{od}\frac{Q}{\Delta E}u(r)=
V_T^{od}\frac{Q}{\Delta E}|(\tilde{nl}S),J':NL,J\rangle
\end{equation}
where Q is a momentum dependent projection operator, $\Delta E(k_1,k_2)$ 
 the energy denominator and $\tilde{nl}$ the correlated two particle
state in the relative coordinates.
In Eq.~(\ref{eq.3}) u(r) is generated as in Ref.~\cite{sha66} by a separation distance
 calculation for the central part of the force in the $^3S_1$ state.
The wave function obtained in this way (full line) heals to the harmonic-oscillator
 wave function (dashed line) as shown in Fig.~1. 
The result obtained for Eq.~(\ref{eq.3}) calculated with the tensor force 
of the Yale potential~\cite{las62}
is given also in Fig.~1 left where we plot  
for the harmonic oscillator size parameter b=1.41~fm:
\begin{equation}
\label{eq.4}
\Psi(\vec{r})=[u(r)Y^1_0(\Omega_{\vec{r}})+w(r)Y^1_2 (\Omega_{\vec{r}})] 
\end{equation}
\begin{minipage}[hb] {0.40\linewidth}
\includegraphics[height=3.4cm]{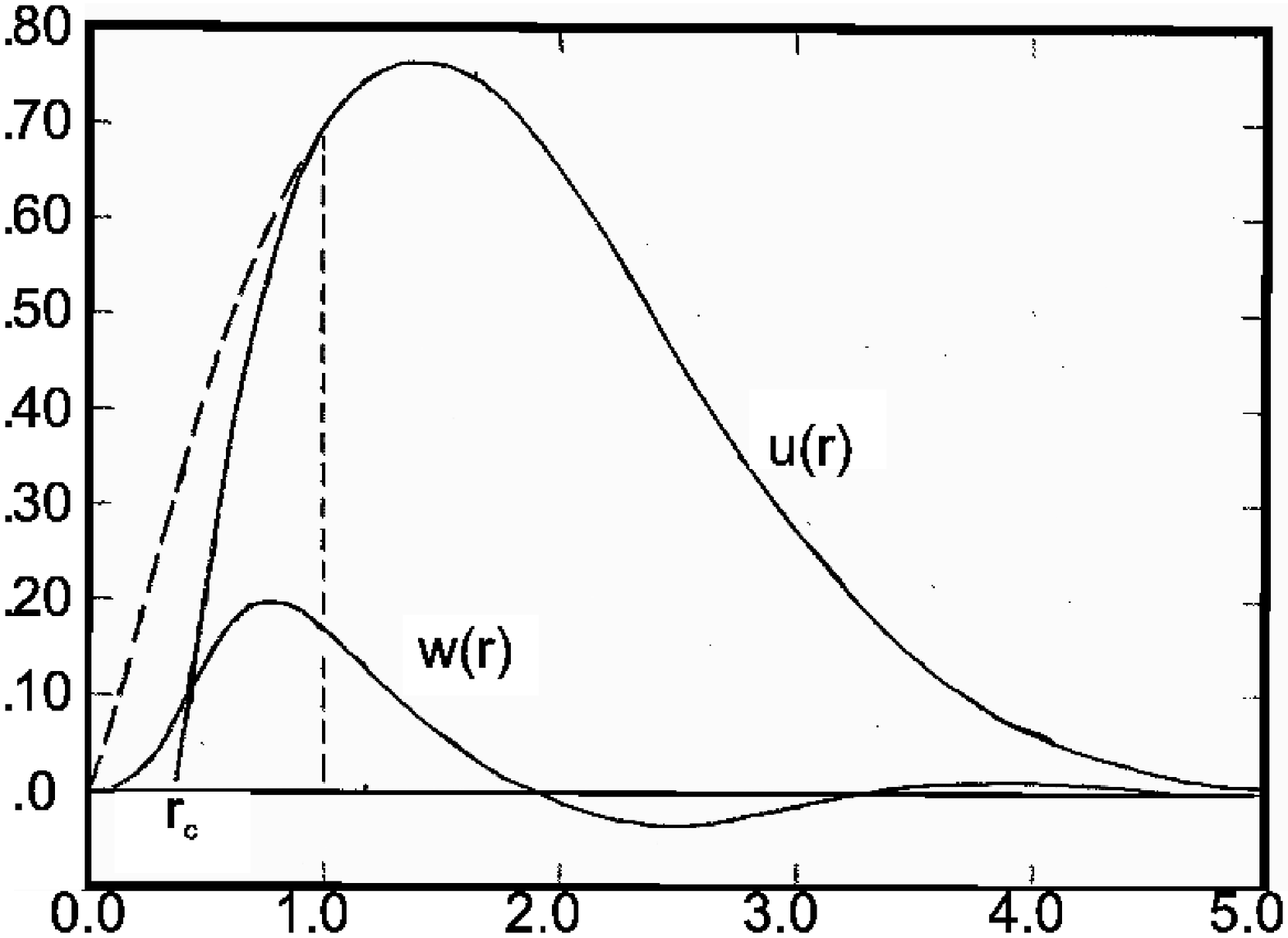}
\end{minipage}
\begin{minipage}[hb]{0.40\linewidth}
\includegraphics[height=3.4cm]{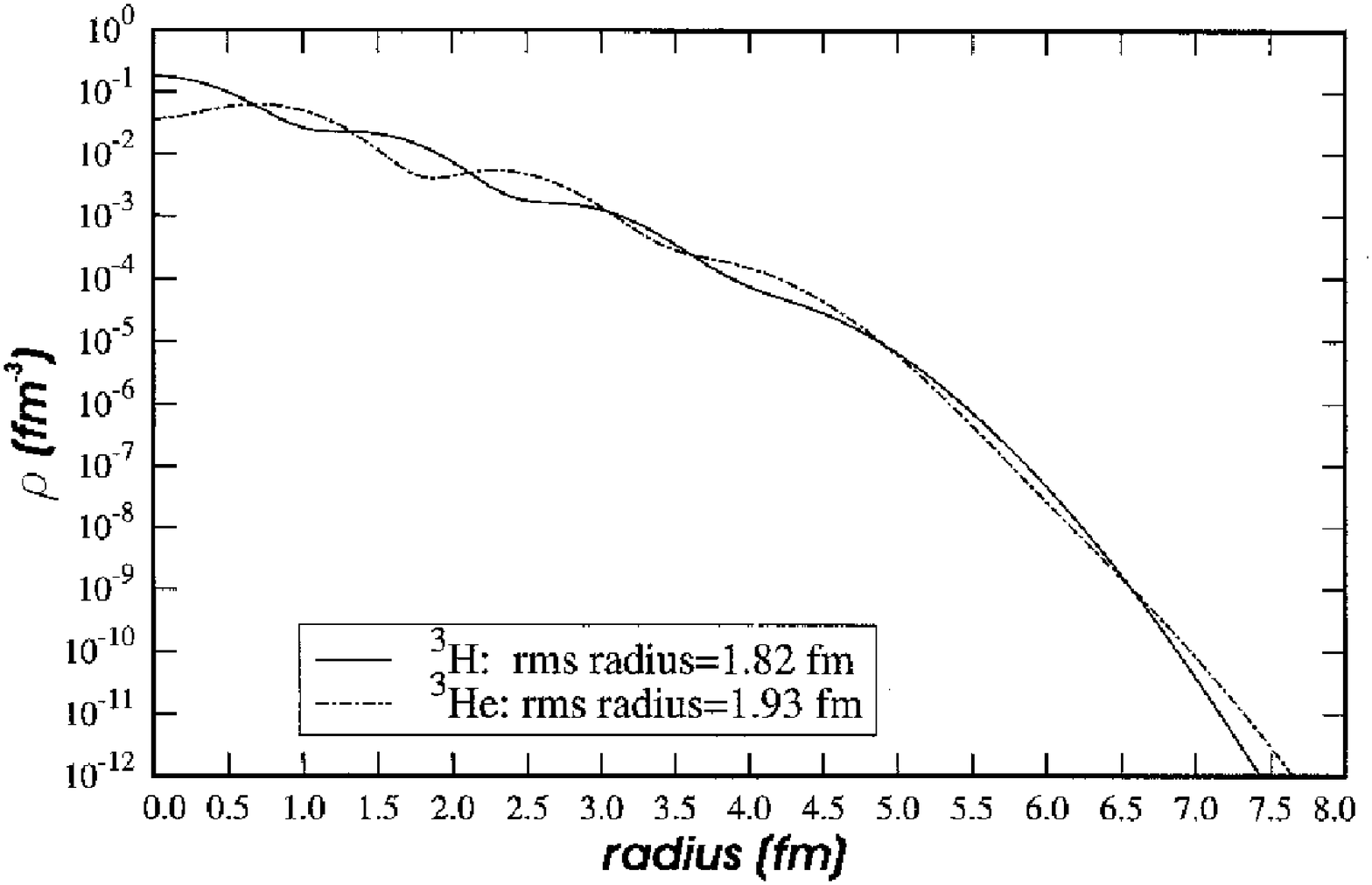}
\end{minipage}
\begin{figure}[htp]
\caption{Left: The u(r) and w(r) wave functions of the deuteron, with
quantum numbers $^3S_1$ and $^3D_1$, plotted as function of r;
Right: Distributions of $^3$H and $^3$He.}
\end{figure}\newline

Being the admixture of the two components, circa $4\%$, 
the wave function Eq.~(\ref{eq.4}) can
be associated to the deuteron wave function.
Let us use then the Hamiltonian Eq.~(\ref{eq.1}) to calculate the structure of
the A=3 nuclei.
Here we propose to calculate the ground state of $^3$H, $^3$He, and $^4$He 
within the EoM
 method which derive the eigenvalue equations by working with the $e^{iS_2}$ 
operator on the wave functions of the A=3,~4 nuclei Ref.
From the diagonalization of the eigenvalue equation of the three particles, 
we obtain an energy difference
$\Delta E(^3H-^3He)$=0.78 MeV and the distributions and radii given in Fig.~1 Right.
 By extending the commutator to a four particle state 
we obtain for the ground state of $^4$He the binding energy of E=28.39 Mev and the rms radius of 1.709 fm.
 In dealing with complex nuclei however the ($S_i,~i=3\cdots n$)
correlations should also be considered.
The evaluation of these diagrams is, due to the 
exponentially increasing number of terms, difficult in a perturbation theory.
We note however that one way to overcome this problem is to work with 
$e^{i(S_1+S_2+S_3+\cdots+S_i)}$ operator on the Slater's determinant
 by keeping the n-body Hamiltonian uncorrelated.
Via the long tail of the nuclear potential the Slater determinant of the 
``n'' particle systems are interacting with the excited Slater's determinants
formed by the (``n'' particles+(mp-mh) mixed-mode excitations). 
The amplitudes of the different determinants are calculated via
the EoM method.
After having performed the diagonalization of the n-body Hamilton's operator we
can calculate the form of the effective Hamiltonian which, by now, includes 
the complete set of the commutator equations.
The method is here applied to $^6$He, $^{11}$Be, $^{14}$C, $^{15}$O, and $^{17}$O. 
A detailed formulation of the model my be found in Ref.~\cite{tom77}. 
\section{Results}
In order to perform structure calculations for complex nuclei, we have to define 
the CMWFs base, the ``single-particle energies'' and to choose the
 nuclear two-body interactions.
The CMWFs are defined as in Ref.~\cite{tom77} by including mixed valence modes
and core-excited states.
The base is then orthonormalized and, since the single particle wave functions
 are harmonic oscillators, the center-of-mass (CM) is removed. 
The single-particle energies of these levels are taken from the known experimental
level spectra of the neighboring nuclei.  
For the particle-particle interaction, we use the G-matrix obtained from
 Yale potential~\cite{sha67}.
These matrix elements are evaluated by applying the
$e^S$ correlation operator, truncated at the second order
term of the expansion, to the harmonic oscillator
base with size parameter b=1.76 fm.   
As elucidate in Refs.~[3] and [4] the effective two-body
 potential used by the DCM and the BDCM models
is separated in low and high momentum components.
 Therefore, the effective model matrix elements calculated within 
 the present separation method
and those calculated by Kuo~\cite{jia01} in the $v_{low-k}$ approximation
 are pretty similar.
The adopted separation method and the $v_{low-k}$ generate two-body
matrix elements which are almost
independent from the radial shape of the different potentials generally
used in structure calculations. \\
The particle-hole matrix elements could be calculated from the particle-particle 
matrix elements via a re-coupling transformation. 
In this contribution we present application of the $S_n$ correlated model to the
charge distributions of $^6$He, $^{11}$Be, and to the electromagnetic transitions
of neutron rich Carbon and Oxygen isotopes. The beta decay strengths from the ground state $^{14}$N
to the excited states of $^{14}$C are also calculated.
In Fig.~2) Left three distributions are given for $^6$He: 
1) the correlated charge distribution  
calculated with the full $S_3$ operator, 
2) the correlated charge distribution  
calculated with the partial $S_3$ operator obtained by neglecting the folded diagrams, 
3) the charge distribution calculated 
for two correlated protons in the $1s_{\frac{1}{2}}$ shell. 
The full $S_3$ correlation operator therefore increases the calculated radii.
In Fig.~2) Right the charge distribution for $^{11}$Be is given.
A charge radius of 3.12 fm has been obtained.
Calculations are performed in a mixed $S_3$ and $S_5$ system. 
The results obtained for the Carbon and Oxygen isotopes are in the 
following presented as function of the increasing valence neutrons.
It is worthwhile to remark that
the high order correlation operators generate the interaction of
the valence particles with the closed shell nucleus.
The correlation model treats therefore consistently the ``A'' particles of
the isotopes.

\begin{center}
\begin{minipage}[hb]{0.40\linewidth}
\includegraphics[height=3.4cm]{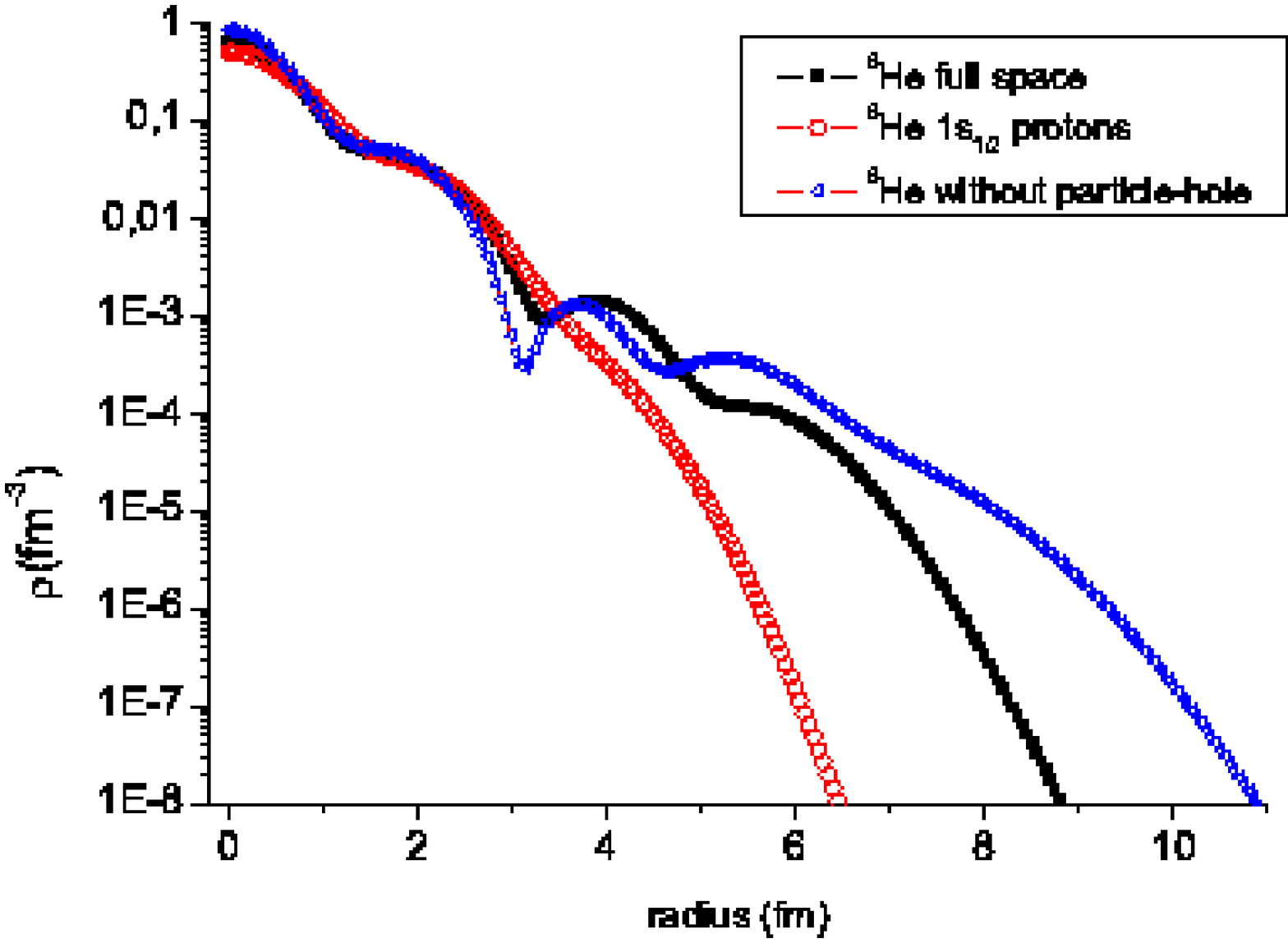}
\end{minipage}
\begin{minipage}[hb]{0.40\linewidth}
\includegraphics[height=3.4cm]{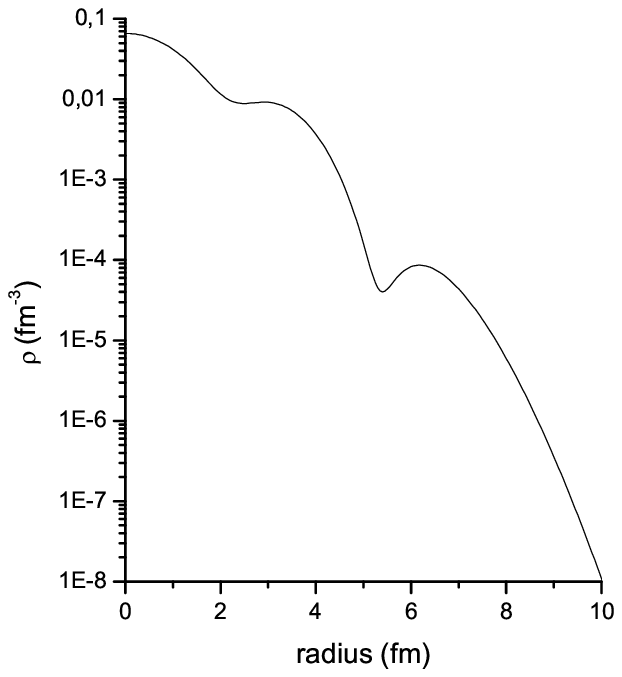}   
\end{minipage}\end{center}
\begin{figure}[htp]
\caption{Left:Charge distributions of $^6$He calculated in different approximations;
Right: Charge distribution of $^{11}$Be.}
\end{figure}
By using generalized linearization approximations and cluster factorization
coefficients~\cite{tom06} we can perform exact calculations. 
In following Tables an over all b=1.76 fm has been used.
  
In Table 1,~3) we give the calculated magnetic moments and rms radii for one-hole and 
for one-particle in $^{16}$O.
 The energy splitting between the ground- and the second (first)
excited states and the electromagnetic transitions for the two isotopes are 
given in Tables 2,~4).
\begin{center}
{\scriptsize \begin{tabular}{c} 
Table~1: Magnetic moment (nm) and rms (fm) of the ground state \\
of $^{15}$O with $J=\frac{1}{2}^-;T=\frac{1}{2}$
\end{tabular}}\end{center}
\begin{center}
{\scriptsize \begin{tabular}{lcc}
                          & DCM & Exp.~\cite{tun06}\\\hline
  Magnetic Moment (mm)    & .70 & .7189 \\ \hline
                          & DCM & Exp.~\cite{dej74}\\ \hline
  rms (fm)                & 2.74& 2.73(3) \\  
\end{tabular}}\end{center}
\begin{center}
{\scriptsize \begin{tabular}{c} Table~2: 
Energy splitting between the ground and the second excited states\\
and the corresponding electromagnetic transitions for $^{15}$O.
\end{tabular}}\end{center}
\begin{center}
{\scriptsize \begin{tabular}{lccccc}
Energy (MeV) & DCM &  Exp.~\cite{tun06} \\ \hline
$ \Delta E_{\frac{1}{2}^-\frac{5}{2}^+}$ & 5.41& 5.24  \\ \hline 
 Ratio & DCM & Exp.~\cite{tun06} \\ \hline 
$\frac{BE(E3;\frac{5}{2}^+\to\frac{1}{2}^-)}{BE(M2;\frac{5}{2}^+\to\frac{1}{2}^-)}$
& .15   & .10    \\ 
\end{tabular}}\end{center}
\begin{center}
{\scriptsize \begin{tabular}{c} Table~3:
       Magnetic moment (nm) and rms (fm) of the ground state\\ 
of $^{17}$O with $J=\frac{5}{2}^+;T=\frac{1}{2}$
\end{tabular}}\end{center}
\begin{center}
{\scriptsize \begin{tabular}{lcc}\hline
                          & DCM & Exp.~\cite{tun06}\\ \hline
  Magnetic Moment (nm)    & -1.88 & -1.89 \\ \hline
                          & DCM   & Exp.~\cite{dej74}\\ \hline
  rms (fm)                & 2.73  & 2.72(3)\\    
\end{tabular}}\end{center}
\begin{center}
{\scriptsize \begin{tabular}{c} Table~4:  Energy splitting between the ground- and the first
excited \\ states and the $E_2$ transition for $^{17}$O.
\end{tabular}}\end{center}
\begin{center}
{ \scriptsize \begin{tabular}{lcc}\hline
Energy (MeV) & DCM & Exp.~\cite{tun06} \\ \hline 
 $\Delta E_{\frac{1}{2}^+\frac{5}{2}^+}$  & 0.87 & 0.89  \\ \hline 
  Transition($e^2fm^4$)  & DCM & Exp.~\cite{tun06} \\ \hline
 $BE(E2;\frac{1}{2}^+\to\frac{5}{2}^+)$ & 2.10   & 2.18$\pm$0.16    \\          
\end{tabular}}\end{center} 

In Table~5) we give the calculated results for the energy splitting between 
the ground- and the $2^+$ excited state and the 
corresponding electromagnetic transition for the $^{14}$C. 
The commutator equations involve $S_2$ and $S_3$ diagrams.
\begin{center}
{\scriptsize \begin{tabular}{c} Table~5:
Calculated energy splitting and $BE(E2;2^+\to0^+)$ transition for $^{14}$C
\end{tabular}}\end{center}
\begin{center}
{\scriptsize\begin{tabular}{lccc}
Energy (MeV) & Ref.~\cite{fuj05} & BDCM & Exp.~\cite{ram87} \\ \hline
 $\Delta E_{0^+2^+}$             &  &8.38 & 8.32  \\ \hline \hline
 Transition($e^2fm^4$) & Ref.~\cite{fuj05} & BDCM & Exp.~\cite{ram87} \\ \hline
 $BE(E2;2^+\to0^+)$  & 3.38   & 3.65     &  $3.74\pm.50$ \\
\end{tabular}}\end{center}
In Table~6) preliminary results for the calculated reduced transition probabilities from the ground state of $^{14}$N
to the $0^+,1^+,2^+$ excited states of $^{14}$C are given. The calculated strengths reproduce reasonably
 well the experimental values~\cite{neg06}.

\begin{center}
{\scriptsize \begin{tabular}{c} Table~6:
Calculated energies of the low-lying states (MeV) $0^+,1^+,2^+$  of $^{14}$C \\and the associated
 reduced transition probabilities $B(GT)$ \\from the J=$1^+$ T=0 ground state of $^{14}$N.
\end{tabular}}\end{center}
\begin{center}
{ \scriptsize  \begin{tabular}{lccccc}
 $^{14}$N  & $J_i^+$T  & $^{14}$C  &   $J_f^+$T  &   Energy (MeV)  &   B(GT) \\ \hline
           & $1^+$0       &           &    $0^+$1      &    0.0    & 0.06     \\
           &              &           &    $0^+$1      &    7.81   & 0.15     \\
           &              &           &    $1^+$1      &    12.17   & 0.12     \\
           &              &           &    $2^+$1      &    7.38   & 0.42     \\
           &              &           &    $2^+$1      &    8.38   & 0.50     \\
           &              &           &    $2^+$1      &    10.91   & 0.35     \\
\end{tabular}}\end{center}

Good results have been overall obtained for the transitions with a neutron effective charge varying between
0.1- to 0.12-$e_n$.  
\section{Conclusion and Outlook}
In this contribution we have investigated the effect of the microscopic 
correlation operators on the exotic structure of the Carbon and
Oxygen isotopes. The microscopic correlation has been separated in short- 
and long-range correlations according to the definition of Shakin.
The short-range correlation has been used to define the effective Hamiltonian
of the model while the long-range correlation is used to calculate the structures and the 
distributions of exotic nuclei. As given in the work of Shakin, only the 
two-body short-range correlation need to be considered in order to derive 
the effective Hamiltonian especially if the correlation is of very short 
range. For the long range correlation operator the three body 
component is important and should not be neglected. Within the three body 
correlation operator, one introduces in the theory a three body interaction 
which compensates for the use of the genuine three body interaction
of the no-core shell model. Within the $S_2$ effective Hamiltonian, good results have been
obtained for the ground state energies and the distributions of
$^3$H, $^3$He, and $^4$He.
The higher order correlation operators $S=3 \cdots n$ have been used 
to calculate the structure and the electromagnetic transitions
of ground and first excited states for the isotopes of Carbon and Oxygen.
By using generalized linearization approximations and cluster factorization
coefficients we can perform expedite and exact calculations.

\end{document}